%% file: main.tex
\RequirePackage{silence}
\documentclass[reprint,amsmath,amssymb,superscriptaddress,nofootinbib,floatfix,aps,pra]{revtex4-2}

\usepackage{tabularx}
\usepackage{amsmath}
\usepackage{amsthm}
\usepackage{amssymb}
\usepackage{graphicx}
\usepackage{xr-hyper}
\usepackage[colorlinks=true, allcolors=blue]{hyperref}
\usepackage{placeins}
\usepackage{xcolor}
\usepackage{braket}
\usepackage{multirow}
\usepackage{makecell}

\usepackage{tikz}
\usetikzlibrary{arrows.meta, positioning}

\makeatletter
\newcommand*{\addFileDependency}[1]{%
  \typeout{(#1)}
  \@addtofilelist{#1}
  \IfFileExists{#1}{}{\typeout{No file #1.}}
}
\makeatother

\newcommand{\liegroup}[0]{\mathfrak{g}}

\newcommand{\su}[1]{\mathfrak{su}(#1)}

\newcommand{\lieclo}[1]{\langle #1 \rangle_{\text{Lie}}}
\newcommand{\braclo}[1]{\langle #1 \rangle_{[.,.]}}
\newcommand{\ipstar}[0]{i \mathcal{P}^*_N}

\newcommand{\ftwo}[0]{\mathbb{F}_2}
\newcommand{\A}[0]{\mathcal{A}}

\newcommand{\rank}[0]{\text{rank}}
\newcommand{\Pa}[0]{\mathbf{P}}
\newcommand{\Po}[0]{\mathcal{P}}

\begin{document}

\title{An Optimized Construction of Lie Algebra Generator Pools for Variational Quantum Eigensolvers in Chemistry}

\author{Yaromir Viswanathan}
\affiliation{Qubit Pharmaceuticals, Advanced Research Department,75014 Paris, France}
\author{Olivier Adjoua}
\affiliation{LCT, Sorbonne Université, UMR 7616 CNRS, 75005 Paris, France}
\author{César Feniou}
\affiliation{LCT, Sorbonne Université, UMR 7616 CNRS, 75005 Paris, France}
\affiliation{Qubit Pharmaceuticals, Advanced Research Department,75014 Paris, France}
\author{Siwar Badreddine}\email{siwar.badreddine@qubit-pharmaceuticals.com}
\affiliation{Qubit Pharmaceuticals, Advanced Research Department,75014 Paris, France}
\author{Jean-Philip Piquemal}\email{jean-philip.piquemal@sorbonne-universite.fr}
\affiliation{LCT, Sorbonne Université, UMR 7616 CNRS, 75005 Paris, France}
\affiliation{Qubit Pharmaceuticals, Advanced Research Department,75014 Paris, France}

\date{\today}

\newtheorem{definition}{Definition}
\newtheorem{theorem}{Theorem} 
\newtheorem{lemma}{Lemma}   
\newtheorem{corollary}{Corollary}[theorem] 
\newtheorem{proposition}{Proposition} 

\newtheorem{example}{Example}
\newtheorem{remark}{Remark}

\definecolor{darkred}{RGB}{210,0,0} 
\newcommand{\cesar}[1]{\textcolor{darkred}{#1}}

\renewcommand{\thetheorem}{\arabic{theorem}} 

\maketitle

\section*{Abstract}
Lie algebras are essential  mathematical structures used in physics to describe sets of quantum operators. Identifying a minimal set of generators to construct these algebras is a central challenge. The traditional search for such generators relies on greedy construction steps applied to an exponentially growing number of candidate operators, making it computationally intractable. Here we show a general, polynomial-scaling strategy, based on fundamental Lie-algebraic properties, to overcome this bottleneck. We apply this framework to quantum chemistry, specifically to adaptive variational algorithms that simulate molecular ground states. By integrating our mathematically verified generator pools into a batched algorithmic framework, we reduce the required quantum resources and improve convergence for strongly correlated systems. Furthermore, this approach eliminates computational bottlenecks that previously restricted fixed-ansatz non-iterative coupled-cluster methods to small molecules, enabling simulations of complex systems well beyond previous limits. This foundational framework also presents broad applications across quantum computing, including quantum error correction, machine learning, and hardware control.

\section*{Introduction}
The evolution of quantum states is fundamentally described by unitary transformations that form a Lie group. The generators of these evolutions are elements of the corresponding Lie algebra, which can be understood as the tangent space at the identity element of the Lie group when seen as a smooth manifold.

Lie algebras are central in quantum computation, where any transformation can be realized as a sequence of unitary evolutions generated by operators from a specific Lie algebra, typically $\su{2^N}$ for a $N$-qubit system. A key research area involves understanding the generation of these algebras from a minimal set of operators. For instance, when using a set of products of Pauli operators, it is known that a minimum of $2N+1$ elements \cite{Smith2025} in this set is required to generate the full basis of $\su{2^N}$ through nested commutation relations. The span of the Lie closure of this basis is known as the dynamical Lie algebra (DLA), a term used in the context of quantum optimal control \cite{Albertini,Wiersema2024,Goh2025,fuclass,Smith2025,class2,Kazi2025}, and later adopted in quantum computing. Consequently, significant previous work has aimed to investigate the specific properties and applications of these bases. In general, these efforts can be categorized into two themes: foundational classification \cite{Wiersema2024,Goh2025,fuclass,Smith2025,class2,Kazi2025} and practical application in variational quantum algorithms \cite{Shkolnikov2023,Sapova2022,Haidar2025}.

One area of research has focused on fundamental properties and complete classification of these Lie algebraic structures based on their generators. Although some studies examined algebras within the context of specific physical models \cite{Wiersema2024,class2,Kazi2025}, a more general method maps the problem to a graph-theoretical setting \cite{fuclass}, allowing the classification of algebras generated by any arbitrary set of Pauli operators without physical constraints.

A related line of research focuses on the practical use of minimal generating sets for the target Lie algebra. These sets, referred to as \textit{minimal complete pools (MCPs)} \cite{Tang2021}, offer a principled approach to the central challenge of constructing efficient unitary transformations in VQE \cite{izmaylov2020order}. This is particularly relevant for adaptive algorithms such as ADAPT-VQE \cite{Grimsley2019}.
Initial works by \cite{Shkolnikov2023,Tang2021,feniou2025greedy} explored the incorporation of complete pools with ADAPT-VQE where \cite{Shkolnikov2023} later demonstrated that for molecular simulations, one needs \textit{symmetry-adapted} pools that preserve the physical symmetries of the Hamiltonian to ensure convergence. To make this approach more practical, the authors in \cite{Sapova2022} developed an automated, greedy algorithm for constructing these symmetry-adapted pools and extended the framework to qubit-tapered systems. Building upon the construction of a complete symmetry-adapted generator set, \cite{Haidar2025} proposed a non-adaptive approach that effectively addresses the gradient measurement bottleneck of the ADAPT-VQE algorithm. 

It is worth noting that each of these studies relies on a set of candidate operators, generally referred to as an operator pool. When such a pool is constructed to be both mathematically complete and of the absolute minimum size required to generate the associated Lie algebra, it is specifically defined as an MCP,
which, in turn, spans the \textit{DLA}. While the specific goals of these studies differ, 
they all seek to answer fundamental questions concerning this set regarding the minimal size required to achieve universality or completeness, possible algebraic structures it can generate, and its optimal construction and selection to efficiently solve a specific problem.

The present work is focused on answering the third question: the optimal and efficient selection of these generators for a target Lie algebra. As highlighted in some of the above studies \cite{Shkolnikov2023,Sapova2022,Haidar2025,fastclosure}, the traditional approach for generating and verifying a complete pool from an exponentially large number of candidate operators can be quite expensive. While our primary motivation is to apply this approach to quantum algorithms for quantum chemistry like ADAPT-VQE, the framework is general and suggests potential utility in other domains.

 In the context of quantum computation, we consider a generator set of Pauli operators. Our approach leverages the mapping
between Pauli operators and the  $\mathbb{F}_{2}^{2N}$
vector space, where $N$ denotes the number of qubits (for further details on this mapping, see Supplementary Note 7). This mapping provides a  framework to analyze the properties of a generator set without resorting to the computationally expensive, explicit construction of the DLAs basis. As highlighted in some of the mentioned studies, the conventional non-greedy approach is to iteratively compute nested commutators and check the dimension of the resulting algebra which scales poorly and is often intractable for even moderately sized systems.

 To set some context for the reader, in variational quantum chemistry, operator pools are used to efficiently construct  an ansatz. These pools are complete, if their elements generate, under nested commutation relations, a basis of the  Lie algebra associated with the system’s Hamiltonian. In this case, there exists, in principle, a sequence of exponentials of these generators that can represent the exact ground state. Variational quantum algorithms are designed to construct and optimize precisely such an ansatz.

A minimal complete pool (MCP) is a complete pool containing the smallest possible number of such  operators.
For a generic $N$-qubit system, the theoretical minimum size for an MCP is $2N-2$ operators \cite{Shkolnikov2023,Sapova2022}. However, a critical challenge arises when applying these generic MCPs to quantum chemistry and the reason for this failure is the presence of symmetries. Molecular Hamiltonians have inherent symmetries that the ground state must obey. To solve this, a different kind of pool is required for quantum chemistry: a  \textit{symmetry-preserving MCP}. These pools are constructed specifically to respect the different properties of the molecular Hamiltonian and their size can be even smaller than $2N-2$, a property we explore mathematically in this paper. While various methods exist to build such user-defined pools, our work presents a theoretical framework for constructing and applying them within variational quantum eigensolvers for chemistry.
\\
\\

In this work, we address this construction bottleneck by introducing a theoretical framework to accelerate the process. By translating the algebraic properties of the generators into a vector space representation mapping Pauli operators to the $\mathbb{F}_{2}^{2N}$ vector space, we reformulate the problem of verifying completeness of a given set of operators. We show that completeness is efficiently determined via the rank and congruence relations of a matrix of size $\mathcal{O}(N^{2})$ derived from the generator's vector space. This provides a general, efficient, scalable, and provable method to verify the minimality and completeness of the generator set for a target Lie algebra, scaling polynomially in $\mathcal{O}(N^{3})$. We validate this framework in variational quantum algorithms for chemistry, introducing MCP-based Batched-ADAPT-VQE (MB-ADAPT-VQE) to reduce measurement overhead and accelerate convergence. Additionally, we demonstrate that our framework eliminates classical bottlenecks that previously restricted fixed-ansatz methods like NI-DUCC-VQE, enabling the simulation of highly complex systems up to 26 qubits. This mathematical framework extends well beyond chemistry.

\section*{Results}

\subsection*
{Summary of Main Lie algebra Results}
\label{results}

Here we give a brief summary of our main results:
Let $\mathcal{A}$ be the set of Pauli operators. Rather than  analyzing the set $\mathcal{A}$ directly, we map it to an associated binary matrix $\mathbf{\Gamma}_{\mathcal{A}} \in \mathcal{M}_{|\mathcal{A}|\times|\mathcal{A}|}(\ftwo)$ (see Definition~\ref{def::gamma}). This matrix contains sufficient information to verify if $\mathcal{A}$ is an MCP. While a full description is provided in the Main theorem section, it is worth mentioning that the computation of this matrix is computationally cheap. The following is our main theoretical results.

\begin{itemize}
\item \textbf{Result 1 (MCP Congruence Theorem):} We establish a formal mapping between operator pools and binary matrices, proving that pool completeness is invariant under  matrix transformations. This allows for the identification of minimal complete pools (MCPs) by evaluating the equivalence classes of $\mathbf{\Gamma}_{\mathcal{A}}$ matrices.
\item \textbf{Result 2 (Polynomial-Time Verification):} Given bracket independence and a canonical MCP for the target Lie algebra, the congruence theorem reduces verification of minimality and completeness to a rank evaluation of the corresponding $\mathbf{\Gamma}_{\mathcal{A}}$ matrices. This reduces the computational complexity to $\mathcal{O}(N^3)$, where $N$ is the number of qubits.
\end{itemize}

The following table summarizes different approaches from the literature for verifying MCPs, alongside our proposed method. For more details on these existing methods, we refer the reader to Supplementary Note 6.
\begin{table}[htpb]
\scriptsize
\centering
\begin{tabular}{|l|l|c|}
\hline
\thead{Approach} & \thead{completeness \\ Guarantee} & \thead{Computational \\ Cost} \\
\hline
\thead{Full Lie algebra$^{\text{\tiny{\cite{Tang2021,Shkolnikov2023}}}}$ \\ check} & \thead{Yes \\ (provable)} & \thead{$\mathcal{O}(4^N)$ \\ exponential} \\
\hline
\thead{Product group$^{\text{\tiny{\cite{Shkolnikov2023}}}}$ \\ size+Inseparability} & \thead{No \\ (numerically robust)} & \thead{$\mathcal{O}(4^N)$ \\ exponential} \\
\hline
\thead{Greedy pool$^{\text{\tiny{\cite{Sapova2022}}}}$ \\ construction} & \thead{No \\ (numerically robust)} & \thead{$\mathcal{O}(N^3)$ \\ polynomial} \\
\hline
\thead{Our work} & \thead{Yes \\ (provable)} & \thead{$\mathcal{O}(N^3)$ \\ polynomial} \\
\hline
\end{tabular}
\caption{\textbf{Comparison of verification methods.} Summary of different approaches from the literature for verifying Minimal Complete Pools (MCPs), alongside our proposed method.}
\label{ref:tablesim}
\end{table}

We summarize the main notations and nomenclature used throughout the paper in the following list.

\subsection*{Notations}

\begin{table}[h]
\centering
\begin{tabular}{|l|l|}
\hline
\textbf{Notation} & \textbf{Description} \\
\hline
$\mathfrak{g}$ & Lie algebra \\
$\mathbf{A},\mathbf{B},\mathbf{C}$ & Matrices\\
$\mathcal{A},\mathcal{B}$ & Sets\\
$|\mathcal{A}|$ & Cardinality of a set\\
$\mathbb{F}_2$ & \begin{tabular}[c]{@{}l@{}}$:=(\{0,1\},\oplus,\cdot)$ denotes the finite \\ field of order two under\\ addition and multiplication modulo 2\end{tabular} \\
$\mathcal{M}_{N\times N}(\mathbb{F}_2)$ & Set of $N \times N$ binary matrices\\
$\otimes$ & Kronecker product (for arrays) \\
$\circ$ & Function composition \\
\hline
\end{tabular}
\label{tab:notations}
\end{table}

\subsection*{Preliminaries}
\label{Framework}
In this section, we introduce the definitions required to define the $\mathbf{\Gamma}_{\A}$ matrix, and subsequently to prove the main Lie algebra results. 

Let $(\mathfrak{g},[.,.])$ be a Lie algebra, i.e. a vector space over a base field $\mathbb{F}$, equipped with a Lie bracket $[.,.]: \mathfrak{g} \times \mathfrak{g} \mapsto \mathfrak{g} $
\begin{definition}(Adjoint map)
For a fixed element $\alpha \in \mathfrak{g}$, its adjoint map $\text{ad}_\alpha$ is the endomorphism on $\mathfrak{g}$ defined as:
  \begin{equation}
  \begin{aligned}
    ad_\alpha : \mathfrak{g} & \mapsto \mathfrak{g} \\
      \beta & \mapsto [\alpha,\beta] := \alpha\beta-\beta\alpha.
  \end{aligned}
  \end{equation}
\end{definition}

\begin{definition}[Bracket-closure]
\label{closure}
Given a set $\mathcal{A} \subseteq \mathfrak{g}$, its bracket-closure is defined as:
\begin{equation}
    \langle\mathcal{A}\rangle_{[.,.]} := \bigcup_{n \in \mathbb{N}} \left\{ \text{ad}_{\alpha_n}\circ \cdots \circ \text{ad}_{\alpha_1}(\alpha_0) \in \mathfrak{g} \;\big|\; (\alpha_j)_{j=0}^n \in \mathcal{A}^{n+1} \right\}.
\end{equation}

This defines the set of all possible adjoint map sequences (or nested commutators) of elements in $\mathcal{A}$. By convention, the set in the union for  $n=0$ is simply $\mathcal{A}$ itself. 
Furthermore, we note that taking the union over all natural numbers $n$ spans all possible combinations of elements from $\mathcal{A}$, therefore, the resulting set is entirely independent of the specific choice of the initial element $\alpha_0$. 
\end{definition}

\begin{definition}(Completeness)
\label{compelteness}
  A complete set for a Lie algebra $\liegroup$ is a subset $\A \ \subseteq  ~\liegroup$ such that 
  \begin{equation}
    \lieclo{\A} := \text{span}_{\mathbb{F}}(\braclo{\A}) = \liegroup.
  \end{equation}
  
                      This defines a set whose Lie-closure is $\liegroup$, or, equivalently, a set whose bracket-closure contains an $\mathbb{F}$-Hamel basis for $\liegroup$ seen as a vector space over $\mathbb{F}$. In practice, it is common to see $\mathbb{F}:=\mathbb{R}$. $\lieclo{\A}$ is also known as the DLA.
  \\
\end{definition}

\begin{definition}(Minimal complete pool (MCP))
  A minimal complete pool (MCP) of a given Lie algebra is a complete set of minimal cardinality in the sense that any other complete set for the Lie algebra will have an equal or larger cardinality.
\end{definition}

In this paper, we will only consider Lie algebras which have a natural basis consisting of Pauli strings. In particular, we will often consider $\su{2^N}$, and its Lie-subalgebras. The Pauli strings can be defined as follows:
\begin{definition} (Pauli Strings) 
  Pauli matrices are a set of $2\times 2$ complex matrices, defined as:
  \begin{equation}
    \begin{aligned}
      &\mathbf{I} := \begin{pmatrix}
      1 & 0 \\
      0 & 1
    \end{pmatrix} \quad \mathbf{X} := \begin{pmatrix}
      0 & 1 \\
      1 & 0 
    \end{pmatrix}\\
    &
    \mathbf{Y} := \begin{pmatrix}
      0 & -i \\
      i & 0 
    \end{pmatrix} \quad
    \mathbf{Z} := \begin{pmatrix}
      1 & 0 \\
      0 & -1
    \end{pmatrix}.
    \end{aligned}
  \end{equation}
A Pauli string (of length $N$) is a matrix of the form $\Pa_1 \otimes \cdots \otimes \Pa_N,\;\Pa_i \in \{\mathbf{I}, \mathbf{X}, \mathbf{Y}, \mathbf{Z}\},~i \in {\{1,\ldots,N\}}$. By convention, the Kronecker product symbol $\otimes$ is suppressed, and Pauli strings are defined as a direct concatenation of $I, X, Y, Z$, where $X, Y, Z$ are short forms for the single-qubit Pauli matrices $\mathbf{X}, \mathbf{Y}, \mathbf{Z}$, respectively. 
In this paper, we may also write Pauli strings by suppressing $\mathbf{I}$, and using subscript indices to denote on which qubit a Pauli matrix acts, e.g. we may write $X_1Y_2Z_4 $ to mean $XYIZ$.
\end{definition}

\begin{definition}(Pauli basis)
\label{pauli_basis}
  The set of length-$N$ Pauli strings, denoted by $\Po_N= \left\{\mathbf{I}, \mathbf{X},\mathbf{Y},\mathbf{Z} \right\}^{\otimes N}$, forms a basis for the Hermitian $2^N \times 2^N$ matrices. We define $\Po_N^*=\Po_N \setminus \left\{ \mathbf{I}^{\otimes N} \right\}$.
\end{definition}
\begin{remark}
 The set $i\Po_N^*$ is a basis for $\su{2^N}$ (the set of skew-Hermitian matrices with vanishing trace). 
The Lie algebras we will consider will always be endowed with the commutator as their Lie bracket. 
\end{remark}
\begin{proposition} (Commutators)
\label{prop_com}
  Let $P_1,P_2 \in i\Po_N^* $, we have
  \begin{equation}
    [P_1,P_2] = \left\{ \begin{array}{cl}
0, & \\
2P_1P_2 = \pm 2P_3, & P_3 \in i\Po_N^*.
\end{array} \right.
  \end{equation}  
Pauli strings can only commute or anti-commute, and in the case that their commutator is nonzero, its result is a Pauli string multiplied by $\pm2$.
\end{proposition}

The proof of this proposition is given in the Supplementary Note 1.

\subsection*{Main theorem}
\label{Methods}
The paper \cite{fuclass} introduces the concepts of an anti-commutation graph and a contraction which we redefine in the following.

\begin{definition}(Anti-commutation graph \cite{fuclass} )
    Given a set of Pauli operators $\A=\{P_i\}_{i\in \{1,\ldots,|\A|\}}$, its anti-commutation graph is defined as $G=(\mathcal{V},\mathcal{E})$, where every vertex corresponds to some $P_i$,
and two vertices are connected by an edge if their corresponding
Paulis anti-commute, i.e.,
\begin{equation}
    \mathcal{V}=\{P_i\}_{i\in \{1,\ldots,|\A|\}}, \mathcal{E}=\{(P_i,P_j)|[P_i,P_j] \neq 0\}.
\end{equation}
\end{definition}

The $\mathbf{\Gamma}_\mathcal{A}$ matrices introduced earlier are precisely the adjacency matrices of these anti-commutation graphs.

\begin{definition}  
\label{def::gamma} 
  Given a set of Pauli strings ${\A := \{\alpha_i\}_{i\in \{1, \cdots, |\A|\}}}$, we define $\mathbf{\Gamma}_\A \in \mathcal{M}_{|\A|\times|\A|}(\mathbb{F}_2)$ as the binary matrix 
  \begin{align*}
    \mathbf{\Gamma}_\A(i,j) := \begin{cases}
    0, \quad [\alpha_i, \alpha_j] = 0 \\
    1, \quad \text{else}.
    \end{cases}
  \end{align*}
  This matrix contains all commutation relations between the elements of $\A$. 
  Note that $\mathbf{\Gamma}_\A$ matrices always have zero diagonal entries and are symmetric.
\end{definition}

\begin{definition}(Contraction \cite{fuclass}) 
  For a set of unsigned Pauli strings $\mathcal{A}$, a contraction of $\beta \in \mathcal{A}$ onto $\alpha \in \mathcal{A}$ (where $\alpha$ and $\beta$ anti-commute) is the process of removing the Pauli string $\alpha$ from $\mathcal{A}$ and replacing it with the unsigned Pauli string that is proportional to their product. 
  Specifically, we replace $\alpha$ with the unsigned string $\gamma \in \{I, X, Y, Z\}^{\otimes N}$ obtained by computing the commutator $[\alpha, \beta]$ and discarding any resulting scalar (e.g., $\pm 2$ or $\pm 2i$).  Formally, when working with Pauli strings in $i\mathcal{P}_N^*$, the contraction is defined as $c_\beta(\alpha)=\pm\tfrac{1}{2}[\alpha,\beta]$, where the sign is chosen such that the result of the contraction yields a valid Pauli string again in $i\mathcal{P}_N^*$. We often chose to implicitly absorb this sign for the sake of simplicity.
\end{definition}

Contracting an anti-commuting Pauli string onto another leaves the Lie closure unchanged, i.e., the new set $\A'$ as in the previous definition will satisfy $\lieclo{\A'} = \lieclo{\A}$. This equality holds strictly because the Lie closure inherently absorbs the scalar prefactor of 2 (originating from the commutator) into its linear span. However, the matrices $\mathbf{\Gamma}_\A$ and $\mathbf{\Gamma}_{\A'}$ will, in general, not be equal. For clarity, we provide an example of this in Figure~\ref{fig:ds}.
\input{figure_contraction}

\begin{remark}
  We emphasize that from this point forward, any mention of an operation or property pertaining to a matrix over $\mathbb{F}_2$ will always be in the context of $\mathbb{F}_2$-arithmetic, i.e., modulo 2.
  
  Furthermore, unless explicitly stated, sets of Pauli strings will be considered as subsets of $\ipstar$, and not $\mathcal{P}_N$.
\end{remark}

The main idea of \cite{fuclass} lies in taking an anti-commutation graph and applying a certain sequence of contractions to it, yielding a graph whose form is known to generate a specific Lie algebra. The canonical forms of these graphs can be found in \cite{fuclass}. Later, the notation $\mathbf{\Gamma}_{\text{canonical}}$ will be used to refer to the adjacency matrix of a canonical graph. What is most impressive about this result is the fact that any anti-commutation graph can be reduced to one of such canonical graphs. 

What we claim is that this idea can be extended to the $\mathbf{\Gamma}_\A$ matrices using a linear algebraic framework. This perspective is more adapted to our final objective of efficiently and optimally constructing the MCPs for a given Lie algebra. 

For the sake of clarity and brevity, we present our results without proofs. Please refer to the \textit{Supplementary Note 3} for additional details and fully worked-out proofs. 

\begin{proposition}
\label{prop:transformation}
  Let $\A$ be a set of Pauli strings, and let $\mathbf{\Gamma}_\A$ be its corresponding adjacency matrix as defined in Definition \ref{def::gamma}. A contraction of $\alpha_j$ onto  $\alpha_i$, ${i,j \in \{1, \cdots, |\A|\}}$, 
   yields  a new set of Pauli strings, denoted $\A'$ and its associated $\mathbf{\Gamma}_{\mathcal{A}'}$ matrix is equal to 
  $$\mathbf{\Gamma}_{\mathcal{A}'} = \mathbf{E}_{i,j}^\top \mathbf{\Gamma}_{\mathcal{A}} \mathbf{E}_{i,j},$$
    where $\mathbf{E}_{i,j} \in \mathcal{M}_{|A|\times|A|}(\ftwo)$ is the identity matrix with its $i$-th column replaced by the sum of the $i$-th and $j$-th columns. Specifically, for $i < j$, $\mathbf{E}_{i,j}$  is defined as:
  $$\mathbf{E}_{i,j} = (\mathbf{e}_1, \dots, \mathbf{e}_{i-1}, \mathbf{e}_i \oplus \mathbf{e}_j, \mathbf{e}_{i+1}, \dots, \mathbf{e}_{|\mathcal{A}|}),$$
where $\mathbf{e}_k \in \mathbb{F}_2^{|\mathcal{A}|}$ denotes the $k$-th standard basis vector.
Simply put, $\mathbf{\Gamma}_{\A'}$ is the matrix obtained from $\mathbf{\Gamma}_A$ by adding its $j$-th column to the $i$-th, and its $j$-th row to the $i$-th.
  
\end{proposition}
An example of the above proposition is illustrated in Figure~\ref{fig:ds}.

Now, since we have established that contractions in the context of graphs correspond to elementary row operations in the context of matrices, it follows that a sequence of contractions can be seen as a matrix operation. As an example, consider a set of Pauli strings $\A$ and its associated $\mathbf{\Gamma}_\A$ matrix. Applying $n$ contractions to the adjacency graph of $\A$ is equivalent to applying $n$ elementary matrices, as in the previous proposition, to $\mathbf{\Gamma}_\A$, denoted by $\mathbf{E}_1, \cdots, \mathbf{E}_n$. The adjacency matrix after this contraction process is given by 
\begin{align*}
    \mathbf{E}_n^\top \cdots \mathbf{E}_1^\top\mathbf{\Gamma}_\A \mathbf{E}_1 \cdots \mathbf{E}_n = (\mathbf{E}_1 \cdots \mathbf{E}_n)^\top \mathbf{\Gamma}_\A (\mathbf{E}_1 \cdots \mathbf{E}_n)
\end{align*}

Noting that a product of elementary matrices yields an invertible matrix, we can elegantly formalize this sequence of transformations as a matrix congruence relation. We first establish this standard definition over $\mathbb{F}_2$:

\begin{definition}[Matrix Congruence]
\label{def:congruence}
Two symmetric binary matrices $\mathbf{A}$ and $\mathbf{B}$ are said to be congruent over $\mathbb{F}_2$, denoted as $\mathbf{A} \sim \mathbf{B}$, if there exists an invertible matrix $\mathbf{P}$ such that $\mathbf{B} = \mathbf{P}^\top \mathbf{A} \mathbf{P} \pmod{2}$.
\end{definition}

Applying this algebraic framework to our anti-commutation graphs, we obtain the following proposition:

\begin{proposition}
\label{propo:trans2}
  Let $\A$ be an arbitrary set of Pauli strings, and let $\mathbf{\Gamma}_\A$ be its corresponding adjacency matrix. Then $\mathbf{\Gamma}_\A$ is congruent to a canonical matrix $\mathbf{\Gamma}_{\text{canonical}}$ over $\ftwo$. That is, there exists an invertible matrix $\mathbf{P} \in \mathcal{M}_{|\A|\times|\A|}(\ftwo)$ such that:
  \begin{equation}
    \mathbf{\Gamma}_{\text{canonical}} = \mathbf{P}^\top \mathbf{\Gamma}_\A \mathbf{P} \pmod{2}.
  \end{equation}
\end{proposition}

The proofs for Propositions~\ref{prop:transformation} and \ref{propo:trans2} are detailed in the Supplementary  Note 3.  in  Proposition~8.

What is left to do now is to find canonical representatives for MCPs of different Lie algebras. Although \cite{fuclass} proposes several families of graphs for their classification which we could transcribe into an adjacency matrix form, we propose an alternate means for classification. The 2008 paper \cite{article} presents the congruence classes for symmetric matrices over $\mathbb{F}_2$. One such family of classes pertains to symmetric matrices with zero diagonal entries, which is precisely what the $\mathbf{\Gamma}_\A$ matrices are. For such matrices, the congruence depends solely on the rank of the matrix. Furthermore, the ranks of such matrices are strictly even. We use this complete classification of symmetric binary matrices of zero diagonal in order to classify MCPs solely based on the rank of their corresponding $\mathbf{\Gamma}_\A$ matrices.

Moving forward, before stating the main theorem, it is necessary to define the crucial concept of a bracket-independent set.
\begin{definition}[\textit{Bracket-independence}]
\label{def::brack}
A set $\mathcal{A}$ of Pauli strings is \textit{bracket-independent} if no element in the set can be expressed as an adjoint sequence of other elements, up to a scalar factor. Formally, for each $\alpha_i \in \mathcal{A}$ and for any scalar $\lambda \in \mathbb{Z}^*$, we have:
  \begin{equation}
    \lambda \cdot \alpha_i \notin \langle \mathcal{A} \setminus \{\alpha_i\} \rangle_{[\cdot,\cdot]},
  \end{equation}
where $\langle \cdot \rangle_{[\cdot,\cdot]}$ is the bracket closure defined in Definition~\ref{closure}. Note that the scalar $\lambda$ must be explicitly accounted for, as nested brackets naturally introduce scaling factors. 
\end{definition}
\begin{remark}
   The inclusion of the integer scalar $\lambda$ is  necessary because the  bracket-closure retains the scaling factors (e.g., powers of 2) generated by successive commutators. Thus, an element is dependent if it can be generated up to an integer scalar. We refer the reader to the Supplementary Note 2, for further details on the notation used in the manuscript on Lie-Closure and bracket-Closure.
\end{remark}

This leads to a fundamental property of MCPs, formalized below.
\begin{proposition} Any MCP set is necessarily bracket-independent. \end{proposition}
The corresponding proofs are given in the Supplementary Note 3 in the proof  of Proposition~10 and Lemmas~1.

With these theoretical foundations established, we now present our main result, Theorem~\ref{theorem:1}, along with its associated corollaries.

\begin{theorem}(MCP Congruence Theorem)
\label{theorem:1}
A bracket-independent set of Pauli strings $\A$ is an MCP for a given Pauli Lie algebra $\mathfrak{g}$ if and only if there exists an invertible matrix $\mathbf{P} \in \mathcal{M}_{|\A|\times|\A|}(\ftwo)$ such that
  \begin{align*}
    \mathbf{P}^\top \mathbf{\Gamma}_{\A} \mathbf{P}= \mathbf{\Gamma}_{\text{canonical-MCP}},
  \end{align*}
  where $\mathbf{\Gamma}_{\text{canonical-MCP}}$ is a canonical matrix for an MCP of $\mathfrak{g}$.
  Equivalently,
  \begin{align*}
  \mathbf{\Gamma}_{\A} \sim \mathbf{\Gamma}_{\text{canonical-MCP}},
  \end{align*}
  where $\sim$ denotes the congruence relation for matrices in $\mathcal{M}_{|A|\times|A|}(\ftwo)$.
\end{theorem}

As previously mentioned, instead of transcribing the canonical graphs presented in \cite{fuclass} to find the $\Gamma_{\text{canonical}}$ matrices, we turn to theorem 1 of \cite{article} which presents a full classification of symmetric binary matrices. The theorem states that if $\mathbf{A}$ is a binary matrix with zero diagonal and rank $k$, then $\mathbf{A}$ is congruent to $\mathbf{J}_k$ and $k$ is necessarily even. Where 

\begin{align*}
    (\mathbf{J}_n)_{ij} := \begin{cases}
        1  & \quad \text{if } (i-j = 1 \text{ and } i \text{ is even}) \\
          & \quad \text{or } (j-i = 1 \text{ and } j \text{ is even}) \\
        0  & \quad \text{otherwise}.
    \end{cases}
\end{align*}  
If $|\A| > k$, the matrix $\mathbf{A}$ is congruent to:
\begin{align*}
    \begin{pmatrix}
        \mathbf{J}_k & 0 \\
        0 & \mathbf{0}_{|\A|-k}
    \end{pmatrix}
\end{align*}
Thus, these $\mathbf{J}_k$ matrices serve as representatives of congruence classes of symmetric binary matrices of zero diagonal, and the following corollary follows

\begin{corollary}
\label{even}
     Let $\A$ be an MCP for a given Pauli Lie algebra $\liegroup$, the rank of $\mathbf{\Gamma}_\A$ is even.
\end{corollary}

\begin{corollary}(Characterizing $\A$ via Matrix Rank) 
\label{theorem:2}
  Let $\A$ be an MCP for a given Pauli Lie algebra $\liegroup$ and let $\A'$ be a set of bracket-independent Pauli strings with cardinality $|\A'| = |\A|$. We have that: 
  \begin{equation*}
  \begin{aligned}
    \A' \text{ is an MCP for }\liegroup \Longleftrightarrow (\rank(\mathbf{\Gamma}_\A)=\rank(\mathbf{\Gamma}_{\A'})) 
  \end{aligned}
  \end{equation*}
\end{corollary}

\begin{corollary}
  Let $\A$ be an MCP for a given Pauli Lie algebra $\liegroup$ and let $\A'$ be a set of bracket-independent Pauli strings. We have that: 
  \begin{equation*}
  \begin{aligned}
    (\rank(\mathbf{\Gamma}_\A)=\rank(\mathbf{\Gamma}_{\A'})) \Longrightarrow \A' \text{ is a complete pool} for  \liegroup 
  \end{aligned}
  \end{equation*}
\end{corollary}

The proofs of this theorem and its corollaries  are given in the Supplementary Note 3.

\subsection*{Outlook on Applications: Quantum Chemistry on Quantum Computers}
\label{section::discussion}
The theoretical framework established in the preceding section enables an approach to the construction of MCPs.

As mentioned in the introduction, MCPs are fundamental building blocks for applications in variational quantum computing, such as the ones investigated in our numerical results. However, the practical utility of these applications is constrained by a significant precomputational bottleneck in the construction and verification of the MCP sets.
For example, published implementations \cite{Shkolnikov2023,Haidar2025} explicitly construct a \textit{product group}, $G \subseteq i\Po_N^*$, from a set of Pauli strings and, in some cases, its associated DLA basis. This explicit enumeration is exponential in the number of qubits, even though generation of a product group can alternatively be checked in polynomial time through a binary symplectic representation and row reduction. These applications have been restricted to systems of limited size, demonstrating results for up to $N=12$ in \cite{Shkolnikov2023} and $N=14$ in \cite{Haidar2025}.

 Our work overcomes this critical limitation. Given a set of Pauli strings, our method analyzes their intrinsic algebraic properties, as described in Theorem \ref{theorem:1}, rather than the extensive properties of the resulting product group and DLAs. This yields a polynomial scaling.

In what follows, we present a brief description of the specific applications employed to benchmark our framework. For these quantum chemistry simulations, the generated MCPs are initialized using a carefully chosen set of physically motivated "starters." These starters serve as the foundational input for our completeness algorithm, which then systematically augments the set with additional elements until the theoretical completeness condition is fully satisfied.

\noindent\textbf{ADAPT-VQE with MCPs:} Here we focus on using our MCP constructor for the Adaptive Derivative-Assembled
Problem-Tailored  Variational Quantum Eigensolver (ADAPT-VQE) \cite{Grimsley2019}. ADAPT-VQE is a variational algorithm designed for approximating the ground state and the ground state energy of a given Hamiltonian $\mathbf{H} \in \mathbb{R}^{2^N \times 2^N} $.
Here, we are investigating molecular systems that possess time-reversal symmetry which yields a real Hamiltonian matrix. Consequently, finding the ground state reduces to a real eigenvalue problem, meaning the exact ground state  can be chosen to be strictly real.
The approximated ground state is a parametrized function that can be defined as:
\begin{equation}
\label{psi::ansätze_def}
        |\Psi\rangle =\prod_{j=1}^M\exp(i\theta_jU_j)|\Psi_0\rangle,
\end{equation}

\noindent where $M \in \mathbb{N}^*$, and $\boldsymbol{\theta}= \{ \theta_j\}_{j=1}^M$ (with $\theta_j \in \mathbb{R}$) is the set of parameters characterizing the state. The operators $U_j$, for $j\in [M]$, are Hermitian  generators acting as excitation operators on the reference state $|\Psi_0\rangle$.
Note that while our minimal complete pools are constructed as subsets of the anti-Hermitian Lie algebra basis $i\mathcal{P}_{N}^{*}$, the operators $U_j$ used in the ansatz are the corresponding standard Hermitian Pauli strings, obtained by factoring out the imaginary unit.

This algorithm's distinct feature is its iterative approach to constructing this parametrized quantum state. At each iteration, the state is expanded by appending an exponentiated generator, which is carefully selected from  an operator pool.
The general workflow of this algorithm is described in the Supplementary Note 4. There are many types of operator pools that can be used to construct the ADAPT-VQE ansätze, however one significant measurement overhead of the traditional ADAPT-VQE is the need to measure the energy gradient for every operator in the predefined operator pool. Traditional approaches \cite{Shkolnikov2023,Sapova2022,liu2021efficient,Grimsley2019,Tang2021,Yordanov2021} typically rely on operator pools, such as those derived from UCCSD excitations, that scale as $\mathcal{O}(N^4)$, where $N$ is the number of qubits. Because the molecular Hamiltonian also comprises $\mathcal{O}(N^4)$ terms prior to any truncation, evaluating the energy gradients for all candidates in the pool requires measuring their commutators with the Hamiltonian. Consequently, the total number of measurements per iteration naively scales as $\mathcal{O}(N^8)$. Therefore, the idea of restricting the pool size is highly attractive. While recent state-of-the-art measurement protocols have elegantly reduced this gradient evaluation cost to $\mathcal{O}(N^5)$ \cite{anastasiou2023really}, the overall algorithmic resource demand remains high. Furthermore, as gradient evaluation costs decrease, the primary computational bottleneck of ADAPT-VQE shifts to the classical optimization phase \cite{Rama2024}. Recent literature has begun to address this shift, demonstrating that improved subroutines, such as recycling the Hessian matrix, can significantly reduce the classical optimizer's overhead \cite{Rama2024}. Consequently, minimizing the pool size itself and reducing optimization macro-iterations remain paramount for practical hardware implementation

The work of \cite{Shkolnikov2023}, introduces a method to mitigate this cost by carefully constructing the operator pool, such that they use:
\begin{itemize}
  \item MCPs as operator pools that scale linearly, $\mathcal{O}(N)$, which reduces the naive measurement overhead for molecular simulations to $\mathcal{O}(N^5)$.
  \item Symmetry-preserving MCPs where the elements of the set are chosen carefully to conserve all relevant symmetries and properties of the problem, as described in the following.
  \item An initial set with sufficient number of symmetry-preserving elements, also known as physically motivated starters, which will be a starting set for the MCP construction, in order to ensure that the algorithm can effectively begin its optimization from the initial Hartree-Fock state.
\end{itemize}

\noindent\textbf{NI-DUCC-VQE approach:} 
As introduced in \cite{Haidar2025}, the NI-DUCC-VQE (Non-iterative Disentangled Unitary Coupled Cluster) approach belongs to the class of "fixed-structure" ansätze (initialized at the start of the VQE process that remain structurally unchanged, with only their parameters updated during optimization). The  Hermitian generators $U_j$ are derived from Pauli strings generated by MCPs. The MCP scales linearly with the number of qubits. The expressivity of the NI-DUCC-VQE ansätze is ensured by employing a construction comprising $k$ layers, such that:
\begin{equation}
    |\Psi\rangle =\prod_{p=1}^k\left(\prod_{j=1}^M\exp(i\theta_{j}U_j)\right)|\Psi_0\rangle,
\end{equation}
where $M=\mathcal{O}(N)$. Unlike iterative ansätze construction methods, this fixed-structure approach avoids the gradient evaluations in the operator selection step.

In our work, we combine these applications with our  MCP construction framework to process molecular systems of larger size.
\\
\\
\noindent\textbf{Symmetries and properties of the pool:}
Because the molecular systems under consideration possess time-reversal symmetry, in the absence of external magnetic fields or relativistic corrections, their exact ground-state wavefunctions can be chosen to be strictly real. Consequently, for these molecular simulations, we restrict our focus to a real-valued ansatz. This imposes a constraint on the operator pool elements, which are restricted to odd Pauli strings (i.e., odd number of Y operators in the string). The complete set of these generators forms the $\mathfrak{so}(2^N)$ Lie algebra. However, molecular Hamiltonians possess additional inherent symmetries, such as the conservation of particle number, total spin, spin projection, and spatial point-group symmetries (see  Supplementary Note 5 for more details). If the pool generators do not respect these symmetries, gradient-based algorithms like ADAPT-VQE can fail. This failure manifests as vanishing gradients, \textit{symmetry roadblock} \cite{Shkolnikov2023}, preventing the algorithm from starting or converging.

In order to avoid this, the complete pool is typically constructed from an initial set of \textit{starters},  which explains why we have adopted the term \textit{physically-motivated} pools. These operators are specifically chosen to preserve all relevant molecular symmetries. Using robust starters, selected via pre-screening criterion to identify the most significant contributing excitations, has been shown to dramatically improve convergence speed \cite{Haidar2025}. The selection of these starters is described below, with further details provided in the Supplementary Methods.

It is worth mentioning that while the set of operators preserving global molecular symmetries (such as total particle number) forms a closed Lie algebra, the specific subset of state-dependent operators used to initialize our pool does not. For example, in this work, \textit{particle number conservation} refers to the specific selection criterion  where starters are defined relative to the Hartree-Fock reference state. Specifically, valid starters are identified as Pauli strings where the number of terms acting on occupied orbitals (annihilations) strictly equals the number acting on virtual orbitals (creations). As detailed in  the Supplementary Note 5, this specific structural property is not closed under the Lie bracket, the commutator of two such strings, may produce a higher-order excitation that violates this balance. However, enforcing this condition is essential for obtaining non-zero gradients: as also shown in \cite{Shkolnikov2023}, employing a generic pool of globally symmetry-preserving operators often leads to vanishing initial gradients because the molecular Hamiltonian cannot couple the reference state to high-order excitations. This state specific condition identifies the subset of operators  that have non-zero overlap with the Hamiltonian's connecting space, ensuring the algorithm can successfully initialize.

Given that completeness requires the final pool to generate a  closed Lie algebra, the algebra generated from these starters will unavoidably expand to include operators that do not preserve all of the initial symmetries. We have therefore investigated which of these properties form a closed subalgebra. Our analysis indicates that the largest subalgebra respecting these constraints is defined by two specific properties: the odd string requirement and the even flip property (the proof is detailed in the  Supplementary Note 5). These defining properties map directly to the generator screening constraints used in the iterative Qubit Coupled Cluster (iQCC) scheme in \cite{Ryabinkin2020} to yield non-vanishing energy gradients. Thus, the algebraic constraints derived in our framework mathematically formalize the heuristic generator screening employed in iQCC.

Furthermore, because the inclusion of the even-flip symmetry restricts the target space to a strictly smaller subalgebra of $\mathfrak{so}(2^N)$ (see  the Supplementary Note 5), the adjacency matrix $\mathbf{\Gamma}$ of any generating pool must have a rank strictly less than $2N-2$. Since this rank must be an even integer (Corollary~\ref{even}), $2N-4$ is the largest possible even rank below this bound. We numerically verified, for the systems studied here, that pools achieving rank $2N-4$ span the target subalgebra. To test for strict minimality, we conducted a leave-one-out analysis by systematically removing a single arbitrary operator from the pool, reducing the pool's size, and re-evaluating the generated Lie algebra for small molecules. In every case, removing one operator caused the dimension of the generated Lie algebra to collapse. This provides numerical evidence that $2N-4$ is the numerical MCP rank for the studied systems; a general proof for this restricted subalgebra remains open. Ultimately, while this rank theoretically guarantees completeness, the distinction between an MCP and a broader CP is less critical for physical implementations. As we demonstrate in later sections, our practical VQE simulations primarily use complete pools rather than MCPs.

\subsection*{Numerical results}

Here, we compare the simulation results for a broad range of molecular systems.  In most benchmarks, we use our theorem \ref{theorem:1} to mathematically guarantee that our operator pools contain a complete generating core, and then, if necessary, we deliberately increase this minimal core into a larger complete pool to improve the algorithmic convergence in our simulations.


Our benchmarks include both iterative and fixed ansätze strategies. For the iterative case, 
 we introduce a MCP-based Batched-ADAPT-VQE (MB-ADAPT-VQE) implementation, and for the fixed ansätze case, we apply the NI-DUCC-VQE approach \cite{Haidar2025} to larger molecules.

These methods are performed across systems ranging from 12 to 26 qubits, from weak to strongly correlated systems. The specific molecular benchmarks include: LiH ($R=1.3\text{{\AA}}$, 12 qubits), the hydrogen chain H$_6$ ($R=3.0 \text{\AA}$, 12 qubits), H$_8$ ($R=0.8 \text{\AA}$, 16 qubits), and H$_2$O (26 qubits, geometry defined in the  Supplementary Table 1 ). All simulations were performed using our Hyperion statevector emulator \cite{hyperion}. We employed the minimal basis set (STO-3G) for most systems, while the $6$-$31G$ basis set was used specifically for the H$_2$O molecule. 
Our Pauli strings are generated using the Jordan\textendash Wigner mapping \cite{Jordan1928}.
We assess ansätze performance across iterations using four metrics: energy accuracy, CNOT count, function evaluations (including the gradient measurements for ADAPT-VQE approaches), and the number of variational parameters.
\\
\\
\noindent\textbf{MB-ADAPT-VQE (iterative ansätze):}
As a first practical test, we present a performance benchmark of an ADAPT-VQE setup, denoted as MB-ADAPT-VQE. This composite strategy uses our  introduced MCPs construction strategy and adds to the ansätze a batch of $k$ operators per iteration (where $k \in \{1,5,10,20,30 \}$), an approach inspired by the so-called batched-ADAPT-VQE, discussed in \cite{Sapova2022}. We employ the standard ADAPT-VQE with qubit-excitation-based (QEB) pool~\cite{Yordanov2021} as a baseline for comparison with MB-ADAPT-VQE.

\begin{figure*}[p]
  \centering
\includegraphics[width=\textwidth, height=0.9\textheight, keepaspectratio]{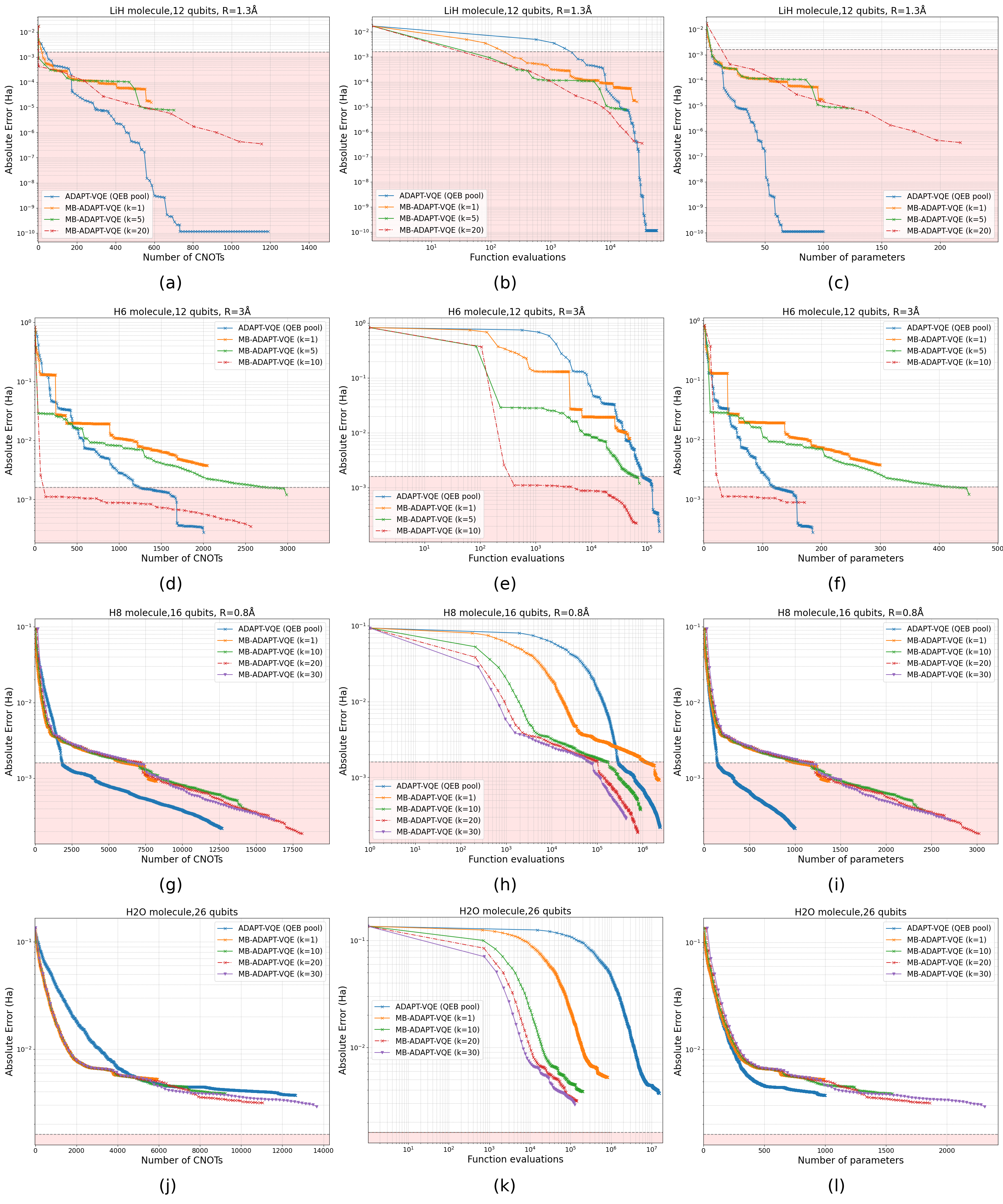}
  \caption{\textbf{Convergence performance of MB-ADAPT-VQE.} Convergence plots for standard ADAPT-VQE (using an $\mathcal{O}(N^4)$ QEB pool) and MB-ADAPT-VQE (using an $\mathcal{O}(N)$ Minimal Complete Pool with varying batch sizes $k$). The rows represent different molecular systems: \textbf{(a-c)} LiH ($R=1.3\text{\AA}$, 12 qubits), \textbf{(d-f)} the hydrogen chain $H_6$ ($R=3.0\text{\AA}$, 12 qubits), \textbf{(g-i)} $H_8$ ($R=0.8\text{\AA}$, 16 qubits), and \textbf{(j-l)} $H_2O$ (26 qubits, geometry defined in Supplementary Table 1). The columns evaluate the absolute energy error against different metrics: number of CNOTs (left column), number of function evaluations (middle column), and number of variational parameters (right column). The red shaded area indicates chemical accuracy (error $< 1.6 \times 10^{-3}$ Ha).}
  \label{Fig::MB}
\end{figure*}

In Figure \ref{Fig::MB}, we compare the convergence of absolute energy errors to the Full-CI (in Ha) versus  function evaluations, number of CNOTs, and number of parameters. The red shaded area represents the chemical accuracy (Absolute Error $< 1.6 \times 10^{-3}$ Ha), the standard ADAPT-VQE (Blue line) uses a QEB pool, and the MB-ADAPT-VQE uses a complete pool generated via the construction strategy introduced earlier.

We first observe a consistent trend across all figures which is the sensitivity of the MB-ADAPT-VQE performance to the batch size $k$. We observe that increasing $k$ generally accelerates the convergence. For instance, in the H$_2$O and H$_8$ systems, curves for $k=20$ and $k=30$ ( red and purple lines) consistently outperform lower batch sizes ($k=1$ or $k=5$), converging faster to the chemical accuracy with significantly fewer function evaluations and comparable CNOT counts than the single-step approach. This suggests that estimating gradients after every individual operator may be unnecessarily costly, particularly for compact operator pools such as in MB-ADAPT-VQE. Similar conclusions have been drawn in recent works \cite{PhysRevResearch.6.013254, Long2024}, which modify how operators are appended to the ansätze by exploring layer-wise subpool exploration strategies to reduce both the circuit depth and measurement costs of VQE.

For small systems (LiH, H$_6$), we observe that using a smaller operator pool is sufficient to reach chemical accuracy. For \textbf{LiH}, standard ADAPT-VQE demonstrates steady convergence and reaches the highest accuracy. In contrast, MB-ADAPT-VQE ($k=20$) exhibits a superior initial convergence rate, crossing the chemical-accuracy threshold significantly faster than the standard method. Lower $k$ values ($k=1, 5$) plateau earlier and struggle to reach the required accuracy. In terms of CNOT count, standard ADAPT-VQE is more efficient, requiring fewer than $60$ parameters and fewer than $600$ CNOTs to reach high precision. While MB-ADAPT-VQE ($k=20$) reaches chemical accuracy quickly in terms of optimization steps and gradient measurements, it incurs a higher parameter count and CNOT count to achieve the same error. For \textbf{H$_6$}, MB-ADAPT-VQE ($k=10$) demonstrates a substantial reduction in measurement overhead, reaching chemical accuracy faster in terms of function evaluations than standard ADAPT-VQE. The standard method shows a long tail requiring more evaluations to cross the chemical-accuracy threshold. MB-ADAPT-VQE ($k=10$) achieves chemical accuracy with fewer than 50 parameters and approximately 200 CNOTs, whereas standard ADAPT-VQE requires nearly 150 parameters and more than 1600 CNOTs to reach the same threshold. For \textbf{H$_8$}, standard ADAPT-VQE regains its advantage. While MB-ADAPT-VQE ($k=20, 30$) performs respectably, standard ADAPT-VQE maintains a steeper slope. Standard ADAPT-VQE reaches chemical accuracy with approximately 100 parameters and $\approx 1{,}500$ CNOTs. In contrast, the MB-ADAPT-VQE variants result in substantially higher parameter and CNOT counts (over 1,000 parameters and more than 8,000 CNOTs) to reach similar error rates. For the larger \textbf{H$_2$O} system,
the standard ADAPT-VQE demonstrates superior parameter efficiency in the early ansätze growth phase ($<500$ parameters), while the MB-ADAPT-VQE strategy sacrifices more parameters and continues to reduce the error significantly further (reaching $\approx 2.5 \times 10^{-3}$ Ha) by using a larger ansätze.

Furthermore, standard ADAPT-VQE has an initially slower convergence phase in terms of function evaluations, whereas MB-ADAPT-VQE ($k=10, 20, 30$) demonstrates a faster error reduction. For example, to reach an error of $10^{-2}$ Ha, MB-ADAPT-VQE ($k=30$) requires approximately two orders of magnitude fewer function evaluations than the standard approach. In terms of CNOT count, Figure~\ref{Fig::MB}j shows that MB-ADAPT-VQE ($k=10, 20, 30$) attains accuracy comparable to standard ADAPT-VQE with fewer CNOT gates.

Further benchmarking of MB-ADAPT-VQE against the recent Coupled Exchange Operator ADAPT-VQE (CEO-ADAPT-VQE), a \cite{Rama2025} framework demonstrates competitive CNOT counts at chemical accuracy and superior function-evaluation scaling (see the Supplementary Discussion).
\\

\begin{figure*}[p]
  \centering
  \includegraphics[width=\textwidth]{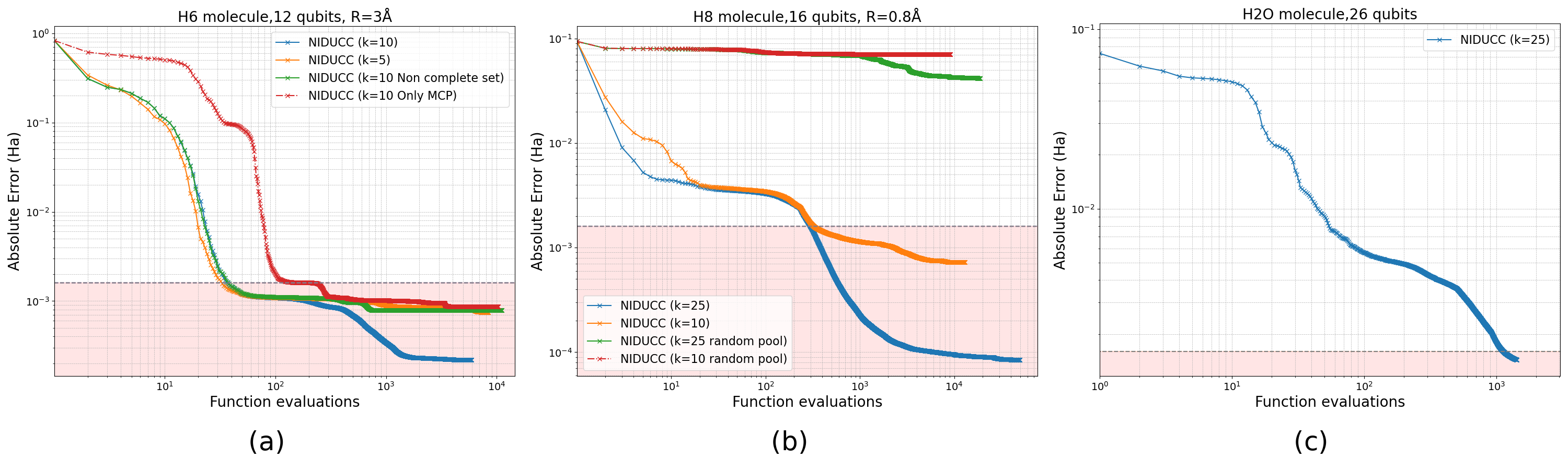}
  \caption{\textbf{Convergence performance of NI-DUCC.} Convergence plots for the NI-DUCC-VQE method displaying absolute energy errors versus the number of function evaluations for varying numbers of layers ($k$). The panels correspond to different molecular systems: \textbf{(a)} stretched hydrogen chain $H_6$ ($R=3.0\text{\AA}$, 12 qubits), \textbf{(b)} $H_8$ ($R=0.8\text{\AA}$, 16 qubits), and \textbf{(c)} $H_2O$ (26 qubits, geometry defined in Supplementary Table 1). In panels \textbf{(a)} and \textbf{(b)}, the convergence of physically-motivated complete pools is compared against incomplete sets, randomly sampled pools, and strictly minimal (MCP-only) pools to highlight the necessity of physics-informed starters. The red shaded area indicates chemical accuracy (error $< 1.6 \times 10^{-3}$ Ha).}
  \label{Fig::niducc}
\end{figure*}

\noindent\textbf{NI-DUCC-VQE (fixed ansätze):}
NI-DUCC-VQE has been previously introduced and already rigorously analyzed  in \cite{Haidar2025}, showing competitive performance against emerging fixed-ansatz methods such as COMPASS \cite{Mondal2023} and COMPACT \cite{Halder2024}. However, prior applications were limited to systems of $N \le 14$ qubits due to the computational bottleneck involved in constructing the minimal complete pool (MCP). Here, we extend the applicability of NI-DUCC-VQE to larger systems and validate our construction strategy across three distinct cases: stretched H$_6$, H$_8$, and H$_2$O. The results, presented in Figure \ref{Fig::niducc}, collectively evaluate convergence efficiency with a focus on the number of layers denoted as $k$. For the strongly correlated system of stretched H$_6$ (12 qubits, $R=3\text{\AA}$), the NI-DUCC-VQE algorithm shows robust convergence, rapidly reaching the chemical accuracy threshold. This performance is achieved with 295 parameters ($1990$ CNOTs) for $k=5$ layers and 590 parameters ($3980$ CNOTs) for $k=10$ layers. Having $k=10$ layers,  proves slightly more efficient, achieving lower final energy errors with a comparable number of function evaluations, a behavior observed in the H$_8$ system as well for ($k=25$, $4350$ parameters, $25800$ CNOTs). Furthermore, for the 26-qubit H$_2$O system, the method achieves chemical accuracy in approximately 1,500 function evaluations with $k=25$ ($16725$ parameters $99850$ CNOTs). This can confirm, the method's viability for larger-scale quantum simulations.

\begin{figure*}[t!]
  \centering
  \includegraphics[width=0.8\textwidth]{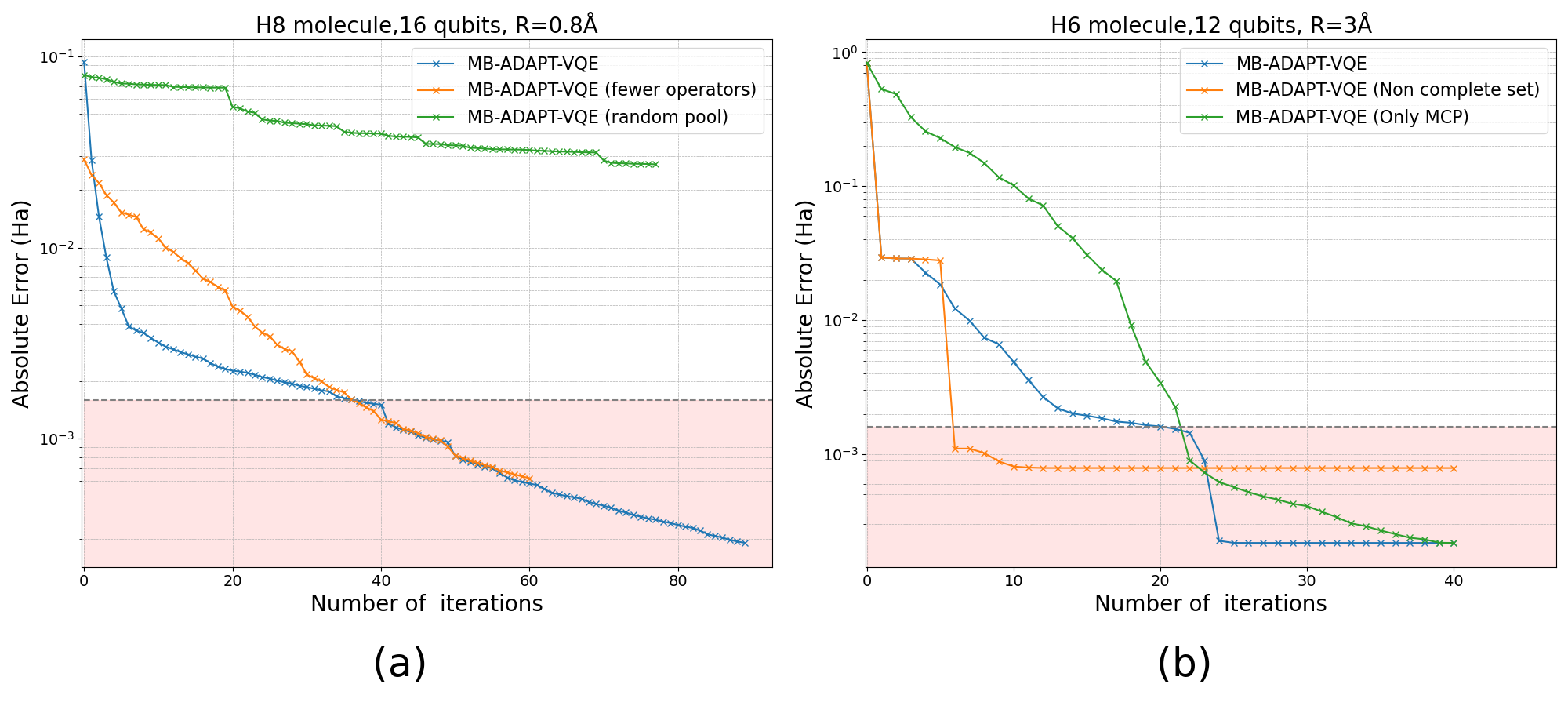}
  \caption{\textbf{Impact of operator pool selection on algorithm convergence.} Absolute energy error versus the number of optimization iterations for the ADAPT-VQE and MB-ADAPT-VQE algorithms. \textbf{(a)} Results for the $H_8$ molecule ($R=0.8\text{\AA}$, 16 qubits) comparing the standard MB-ADAPT-VQE pool against a randomly generated complete pool and a smaller complete pool containing fewer operators. \textbf{(b)} Results for the stretched $H_6$ molecule ($R=3.0\text{\AA}$, 12 qubits) comparing the full MB-ADAPT-VQE pool against an incomplete operator set and a strictly minimal complete pool (Only MCP). The red shaded area indicates chemical accuracy (error $< 1.6 \times 10^{-3}$ Ha).}
  \label{fig::nomcp}
\end{figure*}

\noindent\textbf{Pool choice and completeness:} In Figures \ref{Fig::niducc} and \ref{fig::nomcp},  we evaluate the convergence behavior of MB-ADAPT-VQE and NI-DUCC, focusing specifically on the impact of pool completeness and the selection strategy of the starters.

First, to assess the importance of the initial operator choice, we constructed pools using random selections of single and double excitations. We constructed random candidate operator pools of fixed size $2N-4$ by uniformly sampling from a larger base set. This base set is generated by applying the Jordan\textendash Wigner transformation to fermionic UCCSD single and double excitations, splitting the resulting linear combinations into individual Pauli strings, and removing the $Z$ operators, following standard pool-construction techniques in \cite{Sapova2022}.
We observe in Figure \ref{Fig::niducc}b that the NI-DUCC-VQE approach using these random pools (green and red lines) exhibits a rapid initial plateau at errors between $10^{-2}$ and $10^{-1}$ Ha, failing to reach the chemical accuracy threshold, a similar stagnation is observed with MB-ADAPT-VQE ($k=30$) in Figure \ref{fig::nomcp}a. Second, we show in \ref{Fig::niducc}a that relying only on MCPs with a size of $2N-4=20$ is insufficient. In Figure \ref{Fig::niducc}a, the "Only MCP" in the NI-DUCC-VQE approach ($k=10$) approach (Red line) yields the poorest performance. Although the pool is formally complete, it was recognized early that adding a set of “starter” operators to MCPs in order to diversify search directions, helps avoid early plateaus and improves convergence. Conversely, using physically motivated but non-complete sets (Figures \ref{Fig::niducc}a, \ref{fig::nomcp}b, Orange line) results in a rapid initial descent that stagnates abruptly at the chemical accuracy border. Finally, in Figure \ref{fig::nomcp}a, we analyze a "small" complete pool strategy containing 60 operators (approximately twice the MCP size of 28) instead of $174$, see Table~\ref{ref:tablessizepools}. While it converges more slowly than the standard MB-ADAPT-VQE ($k=20$), it successfully breaks the chemical accuracy threshold with more iterations.\\\\

\noindent\textbf{Benchmarking VQE approaches using MCPs: lessons learned.}
Using a smaller operator pool reduces the number of available update directions at each ADAPT-VQE step, which can lower the likelihood of selecting a high-impact operator at a given iteration. However, it also reduces the computational cost of the operator-selection stage, since ADAPT-VQE requires evaluating an energy gradient for every operator in the pool. MB-ADAPT-VQE is designed to balance this trade-off, reducing function evaluations while preserving the adaptive structure needed to mitigate barren plateaus.
We benchmarked the approach on LiH and H$_8$ molecules, a strongly correlated system (stretched H$_6$), and a larger 26-qubit system (H$_2$O), with performance quantified by function evaluations (including gradient measurements), parameter count, and CNOT gate count. To explicitly quantify the efficiency of our pool construction framework, Table~\ref{ref:tablessizepools} compares the operator counts of our generated pools against standard Qubit Excitation-Based (QEB) pools for various molecular systems. The dimensionality reduction is substantial. For instance, simulating the H$_2$O molecule (26 qubits) typically requires a QEB pool of over 15,000 candidate operators. In  contrast, our algebraic framework guarantees theoretical completeness with a complete pool of just $48$ operators. Furthermore, even after repopulating this minimal set with our physics-informed starters to ensure high expressiveness for the classical optimizer, the final augmented pool (MCP + Starters) contains only $667$ operators. This represents a reduction in pool size of over an order of magnitude compared to standard methods.

The corresponding pool sizes for MB-ADAPT-VQE and for ADAPT-VQE using the QEB pool are given in Table~\ref{ref:tablessizepools} (the exact complete pool configurations are provided in Supplementary Note 8).

\noindent Overall, the numerical results support the expected trends. Standard ADAPT-VQE tends to produce a slightly more compact ansätze, in terms of “operators-added-per-correlation-gained”, for LiH, H$_8$, and to a lesser extent H$_2$O. However, this compactness comes at the cost of a significantly larger number of function evaluations, driven by the size of the operator pool, which ultimately yields a much worse “function-evaluations-per-correlation-gained” ratio compared to MB-ADAPT-VQE.

\noindent Within MB-ADAPT-VQE, increasing the batch size consistently reduces the number of function evaluations required to reach a given accuracy, without degrading ansätze compactness. This establishes batching as an important strategy when working with reduced, but complete, operator pools.

\noindent The behavior also varies across correlation regimes. Standard ADAPT-VQE performs well for weakly correlated and moderately sized systems such as LiH and H$_8$. In contrast, for strongly correlated systems such as stretched H$_6$, MB-ADAPT-VQE with larger batches reaches chemical accuracy more efficiently. For the larger 26-qubit H$_2$O system, both approaches show signs of stagnation, likely due to vanishing gradients, an expected challenge in variational quantum algorithms as system size grows \cite{mcclean2018barren, Larocca_2025} and one that may require additional techniques tailored to large, flat optimization landscapes.

\begin{table}[htpb]
\scriptsize
\begin{tabular}{|l|l|c|c|}
\hline
\thead{Molecule} & \thead{QEB pool size} & \thead{MCP size}  & \thead{MCP+Starters size} \\
\hline
 LiH & 561 & 20 & 34 \\ 
 H$_6$ & 561 & 20 & 59 \\
 H$_8$ & 1940 & 28 & 174 \\
 H$_2$O & 15275 & 48 & 667 \\
\hline

\hline
\end{tabular}
\caption{\textbf{Operator pool dimensionality reduction.} Comparison of the number of elements in standard QEB pools versus our constructed complete pools (MCP + Starters) used in our simulations.}
\label{ref:tablessizepools}
\end{table}

We also benchmarked the performances of the NI-DUCC-VQE approach to extend its applicability to larger systems H$_8$ (16 qubits) and H$_2$O (26 qubits). We have noted that NI-DUCC-VQE exhibits exceptional convergence rates in terms of function evaluations. As shown in Figure \ref{Fig::niducc}, the method reaches chemical accuracy for H$_2$O in approximately $1500$ evaluations. This is a dramatic reduction compared to typical variational runs, making NI-DUCC-VQE highly attractive for reducing the classical optimization overhead. The robustness of this method is further confirmed for H$_8$ and the strongly correlated stretched H$_6$ regime, where increasing the number of layers to $k=10$ ensures convergence to chemical accuracy without being trapped in local minima. The trade-off for this classical speed is a high CNOT count. For H$_2$O, the $k=25$ NI-DUCC-VQE ansatz requires nearly 100,000 CNOTs. While feasible for state-vector emulators or, ultimately, fault-tolerant hardware, such a CNOT count poses significant challenges for current NISQ devices. Recent benchmarks on VQE noise resilience show that the allowable gate-error probability for chemical accuracy scales inversely with the number of entangling gates \cite{Dalton2024}. In light of these entangling-gate and statistical-noise bottlenecks, recent studies \cite{feniou2025greedy} have actively pivoted toward exploring completely gradient-free, local optimization strategies.

Now, turning to the results in Figures \ref{Fig::niducc}a and \ref{Fig::niducc}b, an  observation emerges regarding the choice of operator pools. 
For random complete pools, the rapid stagnation ($10^{-2}$–$10^{-1}$ Ha) observed, shows that the pool requires operators that specifically target the physical system. Furthermore, we have observed that completeness alone does not guarantee efficient convergence and in contrast physically motivated but incomplete pools might fail and suffer from plateaus. Additionally, we have tried a "Small complete pool" strategy (Figure \ref{fig::nomcp}a) which seems to represent a viable middle ground: for example, by increasing the MCP of H$_8$ to ~60 operators instead of having $174$ operators, it reaches the chemical accuracy, but with slower convergence than the full MB-ADAPT-VQE pool. Ultimately, our study confirms that ansatz construction requires physics-informed operator pools, as random or strictly minimal sets lead to suboptimal convergence. Furthermore, it demonstrates that the theoretical completeness criterion alone is insufficient to guarantee an efficient algorithm. Instead, augmenting the minimal pool to a larger size with well-chosen operators allows for significantly faster convergence, a conclusion that aligns directly with the findings of \cite{Sapova2022}.

\section*{Discussion}

In this work, we proposed a general mathematical framework to verify the completeness of a user-defined pool of Pauli string operators with respect to a target Lie algebra. This approach scales polynomially with the number of qubits where this complexity is determined by evaluating the rank of $\mathbf{\Gamma}_\A$ binary matrices derived from the commutation relations of operators in the pool. Unlike known greedy strategies, we demonstrated that algebraic equivalence relationships allow the ranks of these matrices to provide definitive information regarding the completeness of the pool. While this is a general approach applied to a variety of contexts, we specifically validated it within variational quantum algorithms for quantum chemistry.  In this context, we introduced MB-ADAPT-VQE, a batched ADAPT-VQE variant driven by the complete operator pools generated via our method. Furthermore, we  leveraged this framework to construct computationally efficient, highly expressive complete operator pools for fixed-ansatz algorithms such as NI-DUCC.
Because MCPs yield an operator pool that scales as $\mathcal{O}(N)$ rather than the usual $\mathcal{O}(N^4)$, adaptive VQE algorithms benefit from much lower gradient-evaluation overhead. 
However, our numerical investigations, which aligns with   recent findings by Sapova and Fedorov \cite{Sapova2022}, reveal that theoretical completeness alone is insufficient to guarantee an efficient algorithm. Relying on a strictly minimal set often restricts the available optimization directions, leading to suboptimal convergence. Instead, we demonstrate that deliberately increasing this mathematically verified minimal core into a larger complete pool with physically motivated starter operators is practically necessary to navigate the optimization landscape effectively. With these augmented, theoretically complete pools
coupled to batched ADAPT, we achieve faster convergence in challenging regimes with respect to the number of energy evaluations, while also enabling the scalability of the gradient-free NI-DUCC-VQE ansätze previously limited by naive MCP construction cost. It is worth noting that these efforts align with the broad interest in making operator selection more efficient in adaptive variational methods, as seen in recent approaches \cite{huang2025quantum, anastasiou2023really, liu2021efficient, lan2022amplitude}.

Our framework offers a theoretical advantage over greedy, product-group-based construction heuristics, such as those proposed in \cite{Sapova2022}. A primary limitation of the greedy algorithm is its strict reliance on the theoretical size of the target product group as a mandatory, a priori algorithmic input to determine their stopping criterion. Although this characterization is straightforward for the full Pauli space, it becomes  highly non-trivial when imposing  state-specific restrictions on the operator pool. In our work, we construct targeted sub-algebras by strictly enforcing restrictions on the  Pauli operators. While these specific operators are closed under the Lie bracket, calculating the exact theoretical size of the product group corresponding to this heavily restricted subspace is an open  problem. This highlights the core theoretical advantage of our work: our $\Gamma_{\mathcal{A}}$ rank-based theorem entirely circumvents this requirement. Because our framework naively evaluates the underlying Lie-algebraic structure, it mathematically guarantees completeness for these targeted sub-algebras via the rank condition, without needing to search for, or precalculate, the target product group size. Consequently, our methodology provides a fundamentally more adaptable and mathematically robust framework for constructing physically targeted operator pools than existing product-group heuristics.

Overall, this  theoretical framework allows efficient  construction of MCPs for a given Lie algebra. We validated their use in the context of variational quantum algorithms, overcoming limitations of earlier approaches.

Because minimal Pauli generator structures are foundational in stabilizer codes \cite{gottesman1997stabilizer, nielsen00, dauphinais2024stabilizer}, the efficient MCP construction introduced here may support application in quantum error correction, and more generally wherever compact Pauli bases are useful, e.g. in quantum control \cite{li2022quantum}, quantum machine learning and Hamiltonian simulation \cite{patel2024extension}, and might as well contribute to Lie algebras classifications. Concrete avenues for future investigations include:
\begin{itemize}
  \item \textbf{Sub-Lie algebra characterization:} we aim to further characterize the sub-Lie algebras obtained from symmetry-preserving pools. While numerical ranks found for these pools are $2N-4$, the theoretical properties of the sub-algebra generated are still under investigation. Furthermore, the theoretical significance of the ranks of these $\mathbf{\Gamma}_\mathcal{A}$ matrices, and how they can be related to the DLAs, is currently under investigation.
  \item \textbf{Basis selection:} Given the impact of the choice of the operator pool on the convergence of variational algorithms for quantum chemistry, determining the best basis for building MCPs, the number of starters, and the optimal algorithm for their selection remains an open question. Possible ways forward include cheap energy sorting heuristics as introduced in \cite{feniou2025greedy}.
  \item \textbf{Compactification techniques:} We note that MB-ADAPT-VQE lacks a very compact representation that can likely be improved at a cheap cost using overlap-driven compactification techniques as introduced in \cite{feniou2023overlap, feniou2023sparse}. 

Furthermore, while MB-ADAPT-VQE significantly reduces the measurement overhead by compactifying the operator pool, further optimizations can be achieved by modifying how operators are appended to the ansätze. Recent works, such as TETRIS-ADAPT-VQE  \cite{Anastasiou2024} and layer-wise subpool exploration strategies \cite{Long2024}, have demonstrated that circuit depth and measurement costs can be substantially reduced by appending multiple mutually commuting (or spatially disjoint) operators in a single iteration. 
Standard ADAPT-VQE using a UCCSD pool requires evaluating $\mathcal{O}(N^4)$ candidate operators against an $\mathcal{O}(N^4)$ Hamiltonian, via their commutators, leading to a naive $\mathcal{O}(N^8)$ measurement scaling per iteration. Our MCP-based approach reduces the pool size to $\mathcal{O}(N)$, immediately lowering this measurement cost to $\mathcal{O}(N^5)$.
While advanced simultaneous measurement strategies can compress this cost into $\mathcal{O}(N^5)$ commuting sets \cite{anastasiou2023really}, our MCP-based approach  restricts the pool size to  $\mathcal{O}(N)$. By applying these same state-of-the-art grouping techniques to our complete pools, the measurement cost can be lowered to  $\mathcal{O}(N^4)$. One can also combine our compact MCPs with the layer-construction strategies proposed in Refs.~\cite{Anastasiou2024, Long2024} in order to reduce further the measurements cost. This  would reduce the number of required pool-evaluation iterations by an additional factor of $\mathcal{O}(N)$, bringing the overall measurement scaling down to  $\mathcal{O}(N^4)$. We leave the numerical exploration of this highly layer-wise MB-ADAPT-VQE framework to future work.
 It is also important to note that the MB-ADAPT-VQE framework are  complementary to recent algorithmic advancements. As demonstrated in recent literature \cite{Rama2024}, the introduction of  subroutines, such as Hessian recycling, provides  effective methods to reduce the measurement overhead of the classical optimizer in adaptive algorithms. Because MB-ADAPT-VQE shares the same underlying variational and optimization structure, these  same optimization-reduction subroutines can be directly applied to the classical optimizer within the MB-ADAPT-VQE framework to minimize further the total measurement cost.
\end{itemize}

\section*{Methods}

\subsection*{Operator pool construction}
\label{starters}
\begin{figure*}[t]
  \centering
\includegraphics[scale=0.6]{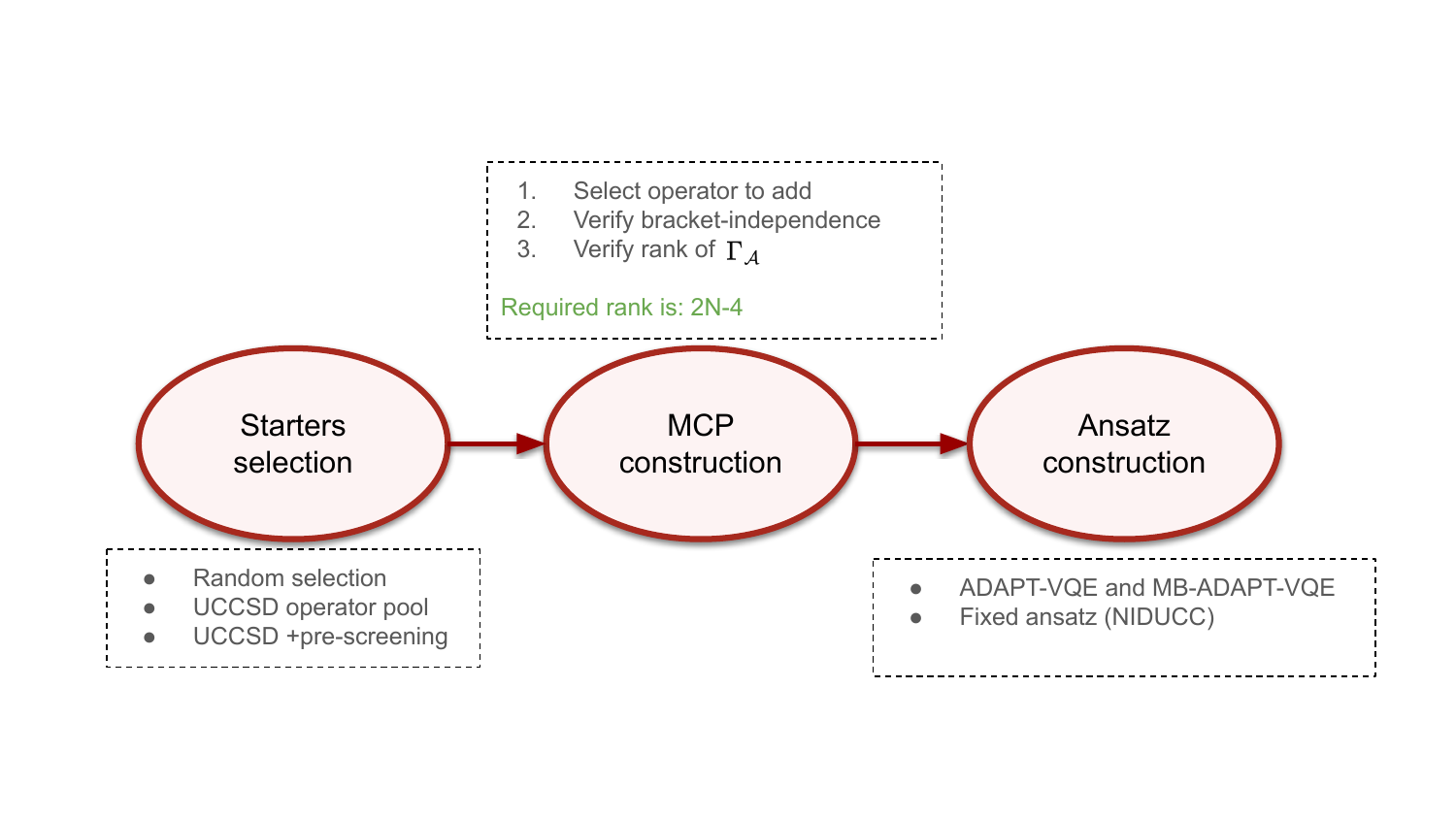}
\caption{\textbf{Operator pool construction workflow.} Schematic detailing the step-by-step generation of a user-defined complete operator pool. The left oval represents the initialization phase, where physically-motivated starters are selected using pre-screening algorithms (e.g., UCCSD amplitudes). The middle oval depicts the core mathematical framework: an operator is added to the pool, verified for strict bracket-independence, and its corresponding adjacency matrix $\Gamma_{\mathcal{A}}$ is evaluated until the target rank of $2N-4$ is achieved to form the Minimal Complete Pool (MCP). The right oval indicates the final integration of this mathematically verified pool into the ansatz construction phase, specifically for execution in algorithms like MB-ADAPT-VQE or fixed-ansatz NI-DUCC.}
  \label{fig:schme}
\end{figure*}

We present an illustrated scheme of the step-by-step construction of a user-defined complete pool in Figure~\ref{fig:schme}. 

For molecular Hamiltonians, the process begins with an initial set of \textit{starters} that must preserve specific physical properties and symmetries (e.g., particle number, spin parity, and spatial point-group symmetry), while consisting of odd Pauli strings with an even number of bit-flips (see the Supplementary Note 5).

Conceptually, these starters could be selected via random sampling from symmetry-preserving subgroups $G \subseteq i\Po_N^*$ or drawn from a standard UCCSD base set. In our implementation, however, we rely on the deterministic pre-screening approach detailed in Ref.~\cite{Haidar2025} and in the  Supplementary Methods. Specifically, this strategy is inspired from the  Unitary Selective coupled-cluster (USCC) method \cite{Fedorov2022,Mitra2024} to select relevant operators by passing them through amplitude thresholds. This strategy begins by generating an initial candidate set comprising all possible double fermionic excitations from the UCCSD ansatz. These excitations undergo a magnitude pre-screening, where they are first filtered based on the magnitude of their two-electron integrals ($|v_{ijkl}| \ge 10^{-2}$). They are subsequently refined through a preliminary VQE run, retaining only the operators where the product of the integral and the optimized amplitude exceeds a defined threshold ($|v_{ijkl} \cdot t_{ijkl}^*| \ge 10^{-3}$), where $t_{ijkl}$ are the amplitudes  corresponding to the excitation operators
already included in the ansatz and determined
by VQE optimization. The surviving fermionic operators are then converted into qubit operators by mapping them to Pauli strings. These operators  are restricted then to odd qubit Pauli strings to guarantee a real wavefunction, and further filtered to  enforce the  symmetries relative to the Hartree-Fock reference state. This physics-informed set of starters initializes our pool construction in all presented simulations, with the explicit exception of Figure 4a, where random selection is deliberately employed as a comparative baseline to demonstrate its algorithmic inferiority.

Our practical pool-construction algorithm then proceeds iteratively from this targeted set of pre-screened starters. First, we systematically extract a bracket-independent subset to initialize the generator pool. We then evaluate the rank of its anti-commutation adjacency matrix, $\mathbf{\Gamma}_\A \in \mathcal{M}_{|\A|\times |\A|}(\ftwo)$, seeking the empirically validated target rank $\operatorname{rank}(\mathbf{\Gamma}_\A)=2N-4$ for the restricted subalgebra studied here. If the initial subset does not reach this rank, the search space is systematically expanded. We first exhaust any remaining symmetry-preserving candidates within the UCCSD pool, similar to the approach in Ref.~\cite{Sapova2022}.
In the worst-case scenario, where the UCCSD pool is exhausted and the rank remains incomplete, our fallback strategy generates candidates on the fly from products of the current pool and symmetry-preserving operators, without explicitly enumerating the entire product group. New elements are validated either by checking linear independence \cite{Sapova2022} or by verifying that they cannot be written as an adjoint sequence. Because the product group grows exponentially, this fallback search has exponential worst-case computational cost. In all numerical benchmarks reported here, the rank $2N-4$ is reached rapidly and explicit construction of the whole product group is avoided.

Once the target rank of $2N-4$ is reached, the core algorithm terminates. For the systems studied here, the rank-$2N-4$ pool behaves as an MCP according to the numerical span and leave-one-out tests described above; a general proof for the restricted subalgebra is left for future work. An MCP alone is often insufficiently expressive to guarantee rapid numerical convergence in VQE optimizations. We therefore repopulate this core with the previously reserved starter elements, yielding a larger complete pool (CP). These augmented CPs are used to construct the ansatz, either via fixed-ansatz methods (e.g., NI-DUCC-VQE) or iterative growth strategies such as ADAPT-VQE and its batched variants \cite{Sapova2022}.

\subsection*{Computational details}
All simulations mentioned above ran onto accelerated computing nodes equipped with NVIDIA A100 and H100 GPU cards. Except for H$_2$O, all other molecules require a single 40GB card from a DGX-A100 machine to host the simulations; while H$_2$O required 8 H100 with 80GB each, due to the storage requirements of the Hamiltonian matrix, i.e., approximately 472 Gigabytes in its Compressed Sparse Row (CSR) format. This latter molecule simulation naturally operated in parallel, and was performed on the Jean Zay supercomputer (IDRIS) at GENCI, whose recent partition is equipped with a set of 4 interconnected H1OO Cards per node, each node exchanging data through a NDR InfiniBand 400GB network.
\section*{Acknowledgments}
GPU computations have been performed at GENCI (IDRIS, Orsay, France) on grant no A0190712052.

\section*{Author contributions}
YV, OA, SB, JPP conceived and designed the research and numerical experiments. Mathematical theorems and proofs were derived by YV with the assistance of SB.  OA and SB performed the numerical experiments. OA, CF, SB, JPP analyzed the data. YV, SB and JPP wrote the paper with the input of all other authors. SB and JPP supervised the work. All authors contributed to reviewing and editing the manuscript.

\section*{Funding}
This work has received funding from the European Research Council (ERC) under the European Union's Horizon 2020 research and innovation program (grant agreement No 810367), project EMC2 (JPP). Support from the PEPR EPIQ - Quantum Software (ANR-22-PETQ-0007, JPP) and HQI (JPP) programs is acknowledged.
\section*{Competing Interests}
JPP is shareholder and co-founder of Qubit Pharmaceuticals. The remaining authors declare no other competing interests. 

\section*{Data availability}
Data generated during the study is available upon request from the authors (E-mail:
jean-philip.piquemal@sorbonne-universite.fr).

\section*{Code availability}

The core C++ and CUDA-accelerated routines for the generation and mathematical verification of the Minimal Complete Pools (MCPs), specifically the algorithms evaluating the   rank condition and verifying strict bracket independence, have been open-sourced. The code is publicly available on GitHub under the MIT License at \url{https://github.com/sbadred/MCP_GAMMA.git}. The full quantum chemistry Variational Quantum Eigensolver (VQE) simulations were performed using the proprietary Hyperion emulator, the integration scripts and extended numerical datasets generated during this study are available from the corresponding authors upon request  (E-mail:
jean-philip.piquemal@sorbonne-universite.fr).

\bibliographystyle{unsrt}
\clearpage
\bibliography{references}

\section*{Figure Captions}

\textbf{Figure 1: Anti-commutation graph simplification.} We analyze the anti-commutation graph corresponding to the set of Pauli strings $\mathcal{A}=\{X_1, Z_1X_2, Z_1Z_2X_3, Z_1Z_2Z_3X_4, Z_4\}$. Black nodes represent the individual Pauli strings, and black edges connect pairs of strings that anti-commute. The red arrow illustrates a contraction operation mapping one node onto another (e.g., contracting $P_1$ onto $P_4$). This operation replaces the target node with their product (resulting in the updated string $P'_4 = Y_1Z_2Z_3X_4$) and updates the graph's connectivity. The corresponding adjacency matrices $\Gamma_{\mathcal{A}}$, $\Gamma_{\mathcal{A}'}$, and $\Gamma_{\mathcal{A}''}$ below explicitly show the row and column additions modulo 2 required at each contraction step.

\textbf{Figure 2: Convergence performance of the batched adaptive variational algorithm.} Convergence plots for standard ADAPT-VQE (using an $\mathcal{O}(N^4)$ QEB pool) and MB-ADAPT-VQE (using an $\mathcal{O}(N)$ Minimal Complete Pool with varying batch sizes $k$). The rows represent different molecular systems: \textbf{(a-c)} LiH ($R=1.3\text{\AA}$, 12 qubits), \textbf{(d-f)} the hydrogen chain $H_6$ ($R=3.0\text{\AA}$, 12 qubits), \textbf{(g-i)} $H_8$ ($R=0.8\text{\AA}$, 16 qubits), and \textbf{(j-l)} $H_2O$ (26 qubits, geometry defined in Supplementary Table 1). The columns evaluate the absolute energy error against different metrics: number of CNOTs (left column), number of function evaluations (middle column), and number of variational parameters (right column). The red shaded area indicates chemical accuracy (error $< 1.6 \times 10^{-3}$ Ha).

\textbf{Figure 3: Convergence performance of the fixed-structure ansatz.} Convergence plots for the NI-DUCC-VQE method displaying absolute energy errors versus the number of function evaluations for varying numbers of layers ($k$). The panels correspond to different molecular systems: \textbf{(a)} stretched hydrogen chain $H_6$ ($R=3.0\text{\AA}$, 12 qubits), \textbf{(b)} $H_8$ ($R=0.8\text{\AA}$, 16 qubits), and \textbf{(c)} $H_2O$ (26 qubits, geometry defined in Supplementary Table 1). In panels \textbf{(a)} and \textbf{(b)}, the convergence of physically-motivated complete pools is compared against incomplete sets, randomly sampled pools, and strictly minimal (MCP-only) pools to highlight the necessity of physics-informed starters. The red shaded area indicates chemical accuracy (error $< 1.6 \times 10^{-3}$ Ha).

\textbf{Figure 4: Impact of operator pool selection on algorithm convergence.} Absolute energy error versus the number of optimization iterations for the ADAPT-VQE and MB-ADAPT-VQE algorithms. \textbf{(a)} Results for the $H_8$ molecule ($R=0.8\text{\AA}$, 16 qubits) comparing the standard MB-ADAPT-VQE pool against a randomly generated complete pool and a smaller complete pool containing fewer operators. \textbf{(b)} Results for the stretched $H_6$ molecule ($R=3.0\text{\AA}$, 12 qubits) comparing the full MB-ADAPT-VQE pool against an incomplete operator set and a strictly minimal complete pool (Only MCP). The red shaded area indicates chemical accuracy (error $< 1.6 \times 10^{-3}$ Ha).

\textbf{Figure 5: Operator pool construction workflow.} Schematic detailing the step-by-step generation of a user-defined complete operator pool. The left oval represents the initialization phase, where physically-motivated starters are selected using pre-screening algorithms (e.g., UCCSD amplitudes). The middle oval depicts the core mathematical framework: an operator is added to the pool, verified for strict bracket-independence, and its corresponding adjacency matrix $\Gamma_{\mathcal{A}}$ is evaluated until the target rank of $2N-4$ is achieved to form the Minimal Complete Pool (MCP). The right oval indicates the final integration of this mathematically verified pool into the ansatz construction phase, specifically for execution in algorithms like MB-ADAPT-VQE or fixed-ansatz NI-DUCC.
\end{document}

%% file: figure_contraction.tex
\begin{figure*}[htpb]
\centering
\resizebox{\textwidth}{!}{
\begin{tikzpicture}[
    every node/.style={align=center},
    v/.style={circle, fill=black, inner sep=2pt},
    thick_red/.style={thick, red!70!black},
    arrow/.style={-{Stealth[scale=1.2]}, thick, red!70!black}
]


\begin{scope}[shift={(0,0)}]
    \node[v, label=left:{$P_1 : X_1$}] (p1) at (-1.5, 0) {};
    \node[v, label=above:{$P_2 : Z_1X_2$}] (p2) at (0, 1.5) {};
    \node[v, label=below:{$P_3 : Z_1Z_2X_3$}] (p3) at (0, -1.5) {};
    \node[v, label={[label distance=0pt]below right:{$P_4 : Z_1Z_2Z_3X_4$}}] (p4) at (1.5, 0) {};
    \node[v, label=right:{$P_5 : Z_4$}] (p5) at (2.5, 1) {};

    \draw[thick, gray] (p1) -- (p2);
    \draw[thick] (p1) -- (p3);
    \draw[thick_red, ->] (p1) -- (p4); 
    \draw[thick] (p2) -- (p3);
    \draw[thick, gray] (p2) -- (p4);
    \draw[thick] (p3) -- (p4);
    \draw[thick] (p4) -- (p5);
\end{scope}

\begin{scope}[shift={(7.5,0)}]
    \node[v, label=left:{$P_1 : X_1$}] (p1) at (-1.5, 0) {};
    \node[v, label=above:{$P_2 : Z_1X_2$}] (p2) at (0, 1.5) {};
    \node[v, label=below:{$P_3 : Z_1Z_2X_3$}] (p3) at (0, -1.5) {};
    \node[v, label={[label distance=0pt]below right:{$P'_4 : Y_1Z_2Z_3X_4$}}] (p4) at (1.5, 0) {};
    \node[v, label=right:{$P_5 : Z_4$}] (p5) at (2.5, 1) {};

    \draw[thick, gray] (p1) -- (p2);
    \draw[thick] (p1) -- (p3);
    \draw[thick] (p1) -- (p4);
    \draw[thick_red, ->] (p3) -- (p2); 
    \draw[thick] (p4) -- (p5);
\end{scope}

\begin{scope}[shift={(15,0)}]
    \node[v, label=left:{$P_1 : X_1$}] (p1) at (-1.5, 0) {};
    \node[v, label=above:{$P'_2 : Y_2X_3$}] (p2) at (0, 1.5) {};
    \node[v, label=below:{$P_3 : Z_1Z_2X_3$}] (p3) at (0, -1.5) {};
    \node[v, label={[label distance=0pt]below right:{$P'_4 : Y_1Z_2Z_3X_4$}}] (p4) at (1.5, 0) {};
    \node[v, label=right:{$P_5 : Z_4$}] (p5) at (2.5, 1) {};

    \draw[thick] (p1) -- (p3);
    \draw[thick, gray] (p1) -- (p4);
    \draw[thick] (p2) -- (p3);
    \draw[thick] (p4) -- (p5);
\end{scope}



\node at (0, -4.5) {
    $\Gamma_{\mathcal{A}} = \bordermatrix{
        & P_1 & P_2 & P_3 & P_4 & P_5 \cr
    P_1 & 0   & 1   & 1   & 1   & 0 \cr
    P_2 & 1   & 0   & 1   & 1   & 0 \cr
    P_3 & 1   & 1   & 0   & 1   & 0 \cr
    P_4 & 1   & 1   & 1   & 0   & 1 \cr
    P_5 & 0   & 0   & 0   & 1   & 0
    }$
};

\node at (7.5, -4.5) {
    $\Gamma_{\mathcal{A}'} = \bordermatrix{
        & P_1 & P_2 & P_3 & P'_4 & P_5 \cr
    P_1  & 0   & 1   & 1   & 1    & 0 \cr
    P_2  & 1   & 0   & 1   & 0    & 0 \cr
    P_3  & 1   & 1   & 0   & 0    & 0 \cr
    P'_4 & 1   & 0   & 0   & 0    & 1 \cr
    P_5  & 0   & 0   & 0   & 1    & 0
    }$
};

\node at (15, -4.5) {
    $\Gamma_{\mathcal{A}''} = \bordermatrix{
        & P_1 & P'_2 & P_3 & P'_4 & P_5 \cr
    P_1  & 0   & 0    & 1   & 1    & 0 \cr
    P'_2 & 0   & 0    & 1   & 0    & 0 \cr
    P_3  & 1   & 1    & 0   & 0    & 0 \cr
    P'_4 & 1   & 0    & 0   & 0    & 1 \cr
    P_5  & 0   & 0    & 0   & 1    & 0
    }$
};

\draw[arrow] (3.25, -4.5) -- (4.25, -4.5);
\draw[arrow] (10.75, -4.5) -- (11.75, -4.5);

\end{tikzpicture}
}
\caption{\textbf{Anti-commutation graph simplification.} We analyze the anti-commutation graph corresponding to the set of Pauli strings $\mathcal{A}=\{X_1, Z_1X_2, Z_1Z_2X_3, Z_1Z_2Z_3X_4, Z_4\}$. Black nodes represent the individual Pauli strings, and black edges connect pairs of strings that anti-commute. The red arrow illustrates a contraction operation mapping one node onto another (e.g., contracting $P_1$ onto $P_4$). This operation replaces the target node with their product (resulting in the updated string $P'_4 = Y_1Z_2Z_3X_4$) and updates the graph's connectivity. The corresponding adjacency matrices $\Gamma_{\mathcal{A}}$, $\Gamma_{\mathcal{A}'}$, and $\Gamma_{\mathcal{A}''}$ below explicitly show the row and column additions modulo 2 required at each contraction step.}
 \label{fig:ds}
\end{figure*}